\documentclass[twocolumn,showpacs,amsmath,amssymb]{jasa}

\usepackage{graphicx}
\usepackage{dcolumn}
\usepackage{bm}

\begin{document}

\title{Elementary Integration Methods for Velocity Excitations in Displacement
Digital Waveguides\footnote{This is a draft, comments and suggestions are very welcome. Please excuse the rough presentation.}}

\author{Georg Essl\footnote{Electronic mail: georg@mle.media.mit.edu}}
\firstauthorlastname{Essl}
\runningtitle{Velocity Excitation Integrators for Waveguides}
 \affiliation{%
Media Lab Europe\\
Sugar House Lane\\
Dublin 8, Ireland}

\received{\today}

\begin{abstract}
Elementary integration methods for waveguides are compared. One using
non-local loading rules based on Heaviside step functions, one using
input side integration and one using integration of the output of
traveling velocity waves. We show that most approaches can be made
consistent with the wave equation in principle, under proper
circumstances. Of all methods studied the Heaviside method is the only
method shown to not suffer from undesirable numerical difficulties and
amendable to standard waveguide loop filtering practices, yet it gives
incorrect results for Neumann boundary conditions. We also discuss
localized velocity excitations, time-limited input-side excitations
and the relation of loop filters to wave variable type.
\end{abstract}

\maketitle

\newpage
\section{Introduction}

The precise meaning of velocity excitations has recently received
renewed interest \cite{Essl04,Smith04}. This interest came out of
comparative use of digital waveguide synthesis \cite{Smith97} and
finite differencing methods \cite{Ames92}. This comparison has
revealed a number of related subtle difficulties that weren't overtly
taken into account in these comparisons. For details see
\cite{Essl04}.

The current paper addresses excitation mechanisms for the purpose of
velocity excitations directly and by example. While basic notions have
long been established \cite{Smith97} defined algorithms and functional
properties haven't been discussed yet.

Specifically comparisons and notions of equivalence usually don't
discuss excitations and their specific implementations. Here we
discuss three basic methods which relate to prior published
non-algorithmic suggestions and unpublished common wisdom in the
field. Overall, published discussions of integration algorithms in
practice are rather rare. Within the field the two sources which are
most explicit about velocity excitations to waveguides are
Smith\cite[or comparable sources by the same author]{Smith97} and
Bank's thesis \cite{Bank00a}.

We will in turn discuss the cases of the infinite string, the string
with fixed boundary conditions, the string with open boundary
conditions, behavior with respect to loop filters, and computational
cost for all three models. We will always use the continuous case as
comparative reference. Then we will discuss interpretive issues with
localized interpretation\cite{Smith97,Smith04}.

\section{Integration methods for the infinite string}

Physically correct behaviors have been discussed by the
author\cite{Essl04}. Related examples have been derived, some
rederived, recently by Smith\cite{Smith04} using a novel state-space
transformation\footnote{This transformation contains many interesting
properties and is highly constructive and hence constitutes a great
recent contribution to the field. For this reason I suggest calling it
the {\em Smith transformation}.}.

\subsection{Heaviside Integration}

There are two related ways to derive the following algorithm. On is to
consider a discretization of the fundamental solution (equation (9) of
the fundamental solution\cite{Essl04}). The other is to find a
discrete implementation of the continuous situation as described by
Smith, citing Morse\cite{Smith97}. The latter will assume an arbitrary
excitation distribution over which one can integrate. The first
assumes impulsive excitations. However any arbitrary excitation
distribution can be seen as the sum of time-shifted impulses, hence
these two are closely related.

The first variant of the algorithm reads as follows: Rescale the
impulse\cite{Smith97}. Add the impulse to all taps left of the
excitation point to the right-going rail of the waveguide. Subtract
(or add with an inverted sign) to all the taps left of the excitation
point to the left-going rail of the waveguide. Repeat for all
excitation positions and impulses at the current time-step.

The second variant of the algorithm reads: Rescale the
distribution\cite{Smith97}. Starting at the right-most position of the
string $x=L$, integrate rescaled excitation distribution to position
$x$ and add the result to the right-going rail and subtract the result
from the left-going rail.

Using either of these algorithms we get for a center excitation
(compare with \cite{Essl04,Smith04}, using Smith's notation
\cite{Smith04}):

\begin{equation}\label{eq:h1}
\begin{tabular}{rrrr|r|rrrrr}
\ldots & \ 1  &\ 1  &\ 1  &\ 1  &\ 0 &\ 0 &\ 0 &\ 0 & \ldots\\
\ldots & -1 & -1 & -1 & -1 &\ 0 &\ 0 &\ 0 &\ 0 & \ldots\\
\hline
\ldots &\ 0  &\ 0  &\ 0  &\ 0  &\ 0 &\ 0 &\ 0 &\ 0 & \ldots\\
\end{tabular}
\end{equation}

\begin{equation}\label{eq:h2}
\begin{tabular}{rrrr|r|rrrrr}
\ldots &\ 1  &\ 1  &\ 1  &\ 1  &\ 1 &\ 0 &\ 0 &\ 0 & \ldots\\
\ldots & -1 & -1 & -1 & \ 0 &\ 0 &\ 0 &\ 0 &\ 0 & \ldots\\
\hline
\ldots &\ 0  &\ 0  &\ 0  &\ 1  &\ 1 &\ 0 &\ 0 &\ 0 & \ldots\\
\end{tabular}
\end{equation}

\begin{equation}\label{eq:h3}
\begin{tabular}{rrrr|r|rrrrr}
\ldots &\ 1  &\ 1  &\ 1  &\ 1  &\ 1 &\ 1 &\ 0 &\ 0 & \ldots\\
\ldots & -1 & -1 & \ 0 & \ 0 &\ 0 &\ 0 &\ 0 &\ 0 & \ldots\\
\hline
\ldots &\ 0  &\ 0  &\ 1  &\ 1  &\ 1 &\ 1 &\ 0 &\ 0 & \ldots\\
\end{tabular}
\end{equation}

The upper two rows are right- and left-going traveling-wave
components. By convention, and following Smith\cite{Smith04}, the
upper rail will move right and the lower rail will move left. The
bottom row is their sum which is the total displacement. The lines
marks the excitation point.

If we observe that we get the correct picture\cite{Essl04,Smith04}. 

We can even revert the direction of propagation and get the
correct time-asymmetric case of a spreading square pulse with negative
sign (compare \cite[Eq. (47)]{Essl04}).

As this solution comes about as the difference of two Heaviside
step-functions, I'll call it the {\em Heaviside integration} method
for waveguides for the unbounded string. We shall see that it readily
extends to the bounded case.

Hence this is one way of loading a waveguide that is physically
accurate (throughout this paper ``physical'' or ``physically
accurate'' will mean, ``comparable results to the continuous solution
of the wave equation''). 

\subsection{Input-side integration}

Very few papers discuss velocity excitations explicitly. Bank
\cite{Bank00a,Bank00b} is an exception. He employs what I will call
``input-side integration''. The idea is to integrate a velocity input
before feeding it into a waveguide to arrive at a velocity
excitation. A procedure that is suggested by the integral relationship
of the two \cite{Smith97}.

If we interpret the waveguide to be a spatial discretization of the
string with both rails sharing the same spatial position and we excite
at this spatial point, we get the following result to an impulsive
excitation:

\begin{equation}
\begin{tabular}{rrrr|r|rrrrr}
\ldots &\ 0  &\ 0  &\ 0  &\ 1  &\ 0 &\ 0 &\ 0 &\ 0 & \ldots\\
\ldots &\ 0  &\ 0 & \ 0 & \ 1  &\ 0 &\ 0 &\ 0 &\ 0 & \ldots\\
\hline
\ldots &\ 0  &\ 0  &\ 0  &\ 2  &\ 0 &\ 0 &\ 0 &\ 0 & \ldots\\
\end{tabular}
\end{equation}

\begin{equation}
\begin{tabular}{rrrr|r|rrrrr}
\ldots &\ 0  &\ 0  &\ 0  &\ 1  &\ 1 &\ 0 &\ 0 &\ 0 & \ldots\\
\ldots &\ 0  &\ 0  &\ 1  &\ 1  &\ 0 &\ 0 &\ 0 &\ 0 & \ldots\\
\hline
\ldots &\ 0  &\ 0  &\ 1  &\ 2  &\ 1 &\ 0 &\ 0 &\ 0 & \ldots\\
\end{tabular}
\end{equation}

\begin{equation}
\begin{tabular}{rrrr|r|rrrrr}
\ldots &\ 0  &\ 0  &\ 0  &\ 1  &\ 1 &\ 1 &\ 0 &\ 0 & \ldots\\
\ldots &\ 0  &\ 1  &\ 1  &\ 1  &\ 0 &\ 0 &\ 0 &\ 0 & \ldots\\
\hline
\ldots &\ 0  &\ 1  &\ 1  &\ 2  &\ 1 &\ 1 &\ 0 &\ 0 & \ldots\\
\end{tabular}
\end{equation}

We see a peak at the excitation point that doesn't exist in the
correct simulation discussed earlier and the continuous
simulation. It's called {\em Bank's anomaly} after Balazs Bank who was
the first to point it out\cite{Bank00a}. This anomaly specifically
appears at the point of excitation, which is exactly how Bank found
it. One can show that it also appears at the center-symmetric position
on the string due to constructive interference. A non-linear hammer
coupling needs to know the local displacement. A question remains to
answered, which is whether the anomaly disappears when the excitation
is completed. But we see an immediate way to resolve it. If the
excitation point is between spatial sampling points, the anomaly
disappears. Hence we cannot naively chose excitations on the spatial
sampling point without taking the anomaly into account. See Bank for a
number of possible resolutions \cite{Bank00a}. These yet lack a clear
physical interpretation. {\em Bank anomaly} points at the difficulty
of spatial representation of excitation points, a topic yet to be
explored in detail. I will not attempt to address it here.

To get the correct time-asymmetric pattern upon inverting the
direction of the reals, we need to invert the signs of the excitation.

\subsection{Output-side Integration}

Finally one can consider ``output-side integration''\footnote{I'm
unaware of this being explicitly introduced elsewhere. If the reader
knows of a prior reference, please let me know.}. Here we integrate the
sum of rails carrying velocity waves to get one accumulated
displacement representation. The following diagram contains a fourth
row, which contains the integration of the sum above it.

\begin{equation}
\begin{tabular}{rrrr|r|rrrrr}
\ldots &\ 0  &\ 0  &\ 0  &\ 1  &\ 0 &\ 0 &\ 0 &\ 0 & \ldots\\
\ldots &\ 0  &\ 0 & \ 0 & \ 1  &\ 0 &\ 0 &\ 0 &\ 0 & \ldots\\
\hline
\ldots &\ 0  &\ 0  &\ 0  &\ 2  &\ 0 &\ 0 &\ 0 &\ 0 & \ldots\\
\hline
\ldots &\ 0  &\ 0  &\ 0  &\ 2  &\ 0 &\ 0 &\ 0 &\ 0 & \ldots\\
\end{tabular}
\end{equation}

\begin{equation}
\begin{tabular}{rrrr|r|rrrrr}
\ldots &\ 0  &\ 0  &\ 0  &\ 0  &\ 1 &\ 0 &\ 0 &\ 0 & \ldots\\
\ldots &\ 0  &\ 0  &\ 1  &\ 0  &\ 0 &\ 0 &\ 0 &\ 0 & \ldots\\
\hline
\ldots &\ 0  &\ 0  &\ 1  &\ 0  & 1 &\ 0 &\ 0 &\ 0 & \ldots\\
\hline
\ldots &\ 0  &\ 0  &\ 1  &\ 2  & 1 &\ 0 &\ 0 &\ 0 & \ldots\\
\end{tabular}
\end{equation}

\begin{equation}
\begin{tabular}{rrrr|r|rrrrr}
\ldots &\ 0  &\ 0  &\ 0  &\ 0  &\ 0 &\ 1 &\ 0 &\ 0 & \ldots\\
\ldots &\ 0  &\ 1  &\ 0  &\ 0  &\ 0 &\ 0 &\ 0 &\ 0 & \ldots\\
\hline
\ldots &\ 0  &\ 1  &\ 0  &\ 0  &\ 0 &\ 1 &\ 0 &\ 0 & \ldots\\
\hline
\ldots &\ 0  &\ 1  &\ 1  &\ 2  &\ 1 &\ 1 &\ 0 &\ 0 & \ldots\\
\end{tabular}
\end{equation}

Hence we see that input-side and output-side integrations behave
comparably. Input-side integration requires only one integrator
whereas the output-side case requires one for each spatial point. If a
read-out point is local, then this can however also reduced to one
integrator. We too see that Bank's anomaly persists for excitations on
a spatial point. It can be resolved by subtracting half the excitation
from the integrator at the excitation point at the moment of
excitation. The output-side integration also suffers from the
numerical weaknesses of non-leaky integrators [Draft note: need
discussion of leaky versus non-leaky
integrators. See\cite{Smith97}.]. In the input-side case this problem
is contained. This we will discuss when introducing boundary
conditions.

These two approaches have, however a crucial difference. The {\em
content of the traveling waves differ} and hence in general the {\em
filters in the loop differ} if they wants to achieve the same final
result\footnote{Of course there is only one physically accurate
result.}. After all in one case the filter will see impulses as input
whereas in the other case it will be step functions\footnote{I am
unaware of any publication that point to the important of the choice
of the wave variable to the properties of loop filters, or
alternatively speaks to the relation of impulse response to damping
read-out. Again, if a reader knows a reference, please let me know. I am aware that Smith discusses the somewhat related issues of observability\cite{Smith03}.}.

\subsection{Hybrid methods}

A number of hybrid approaches have been proposed. I will not attempt
to discuss them here, as the goal is the understand velocity
excitations in a purely waveguide setting. See
for example \cite{Karjalainen03,KE03,KS03}.

\section{Effect of Boundary Conditions}

Next we will introduce boundary conditions. This raises another
question. How do the respective approaches to integration compare with
respect to boundary conditions.

First we refer to \cite{Essl04} for the ground truth. The solution is
periodically extended at boundaries which create mirror images of the
solution (hence the name ``method of images'' in the theory of partial
differential equations \cite{Taylor96}). The image is so chosen as to
satisfy the boundary condition. 

In the case of a string with fixed ends (Dirichlet boundary
conditions) it is well-known that for displacement waves the boundary
inverts the sign, hence the image is sign-inverted with respect to the
original. See Figure 3 in \cite{Essl04} for a center excitation. If
the excitation is off-center we get a parallelogram with maximum width
less than the length of the string. In either case there are three
possible states: unit positive, unit negative and zero extension. The
three states alternate with a zero extension state always between
different sign unit extension ones. [Draft note: Add figure for
off-center case, comparable to Figure 3 in \cite{Essl04} to illustrate
this.]

Also we observe that the transition between the vanishing of negative
to positive extension looks like the discrete case illustrated in Eq.
(47) of \cite{Essl04} and hence corresponds to the case also observed
under time-reversal for the Heaviside integration method.

In the case of a string with open ends (Neumann boundary conditions)
the displacement waves do not invert sign at the boundary. Here we get
a linear accumulation with every reflection. The geometric picture is
the same as the Dirichlet case, except that former zero extension
states have even increasing accumulation, and the other states for
odd increasing accumulation. See also \cite{Essl04} for a formal
derivation of this property.

\section{Integration Methods and Dirichlet Boundary Conditions}

Inverting boundary conditions should create the right image
of traveling waves. We will thus use these conditions and observe the
various methods.

\subsection{Heaviside Integration}

These are the respective results of the Heaviside Integrator for time
steps equivalent to half a string length. The diagram is as before
except that vertical lines at each side denote the boundary. The
excitation is at the midpoint:

\begin{equation}
\begin{tabular}{|rrrr|rrrr|}
1  & 1  & 1  & 1  &\ 0 &\ 0 &\ 0 &\ 0 \\
-1 & -1 & -1 & -1 &\ 0 &\ 0 &\ 0 &\ 0 \\
\hline
\ 0  &\ 0  &\ 0  &\ 0  &\ 0 &\ 0 &\ 0 &\ 0 \\
\end{tabular}
\end{equation}

\begin{equation}
\begin{tabular}{|rrrr|rrrr|}
\ 1  &\ 1  &\ 1  &\ 1  &\ 1 &\ 1 &\ 1 &\ 1 \\
0  &\ 0  & \ 0 & \ 0 &\ 0 &\ 0 &\ 0 &\ 0 \\
\hline
1  &\ 1  &\ 1  &\ 1  &\ 1 &\ 1 &\ 1 &\ 1 \\
\end{tabular}
\end{equation}

\begin{equation}
\begin{tabular}{|rrrr|rrrr|}
0  &\ 0  &\ 0  &\ 0  &\ 1 &\ 1 &\ 1 &\ 1 \\
0  &\ 0  & \ 0 & \ 0 &-1 &-1 &-1 &-1 \\
\hline
0  &\ 0  &\ 0  &\ 0  &\ 0 &\ 0 &\ 0 &\ 0 \\
\end{tabular}
\end{equation}

\begin{equation}
\begin{tabular}{|rrrr|rrrr|}
0  &\ 0  &\ 0  &\ 0  &\ 0 &\ 0 &\ 0 &\ 0 \\
-1  &-1  & -1 & -1 &-1 &-1 &-1 &-1 \\
\hline
-1  &-1  &-1  &-1  &-1 &-1 &-1 &-1 \\
\end{tabular}
\end{equation}

Observe that the pattern repeats and matches the continuous case. The
off-center case can readily be plugged in for similar results.

\subsection{Input-side Integration}

The excitation is placed in the middle of the string and constitutes
loading the result of a non-leaky integrator fed by an impulse to
equal parts left and right of the excitation-point into the respective
traveling waves.

\begin{equation}
\begin{tabular}{|rrrr|rrrr|}
0  &\ 0  &\ 0  &\ 0  &\ 0 &\ 0 &\ 0 &\ 0\\
\ 0  &\ 0 & \ 0 & \ 0  &\ 0 &\ 0 &\ 0 &\ 0 \\
\hline
\ 0  &\ 0  &\ 0  &\ 0  &\ 0 &\ 0 &\ 0 &\ 0 \\
\end{tabular}
\end{equation}

\begin{equation}
\begin{tabular}{|rrrr|rrrr|}
\ 0  &\ 0  &\ 0  &\ 0  &\ 1 &\ 1 &\ 1 &\ 1 \\
\ 1  &\ 1  &\ 1  &\ 1  &\ 0 &\ 0 &\ 0 &\ 0 \\
\hline
\ 1  &\ 1  &\ 1  &\ 1  &\ 1 &\ 1 &\ 1 &\ 1 \\
\end{tabular}
\end{equation}

\begin{equation}
\begin{tabular}{|rrrr|rrrr|}
-1  &-1  &-1  &-1  &\ 1 &\ 1 &\ 1 &\ 1 \\
\ 1  &\ 1  & \ 1 &\ 1  &-1 &-1 &-1 &-1 \\
\hline
\ 0  &\ 0  &\ 0  &\ 0  &\ 0 &\ 0 &\ 0 &\ 0 \\
\end{tabular}
\end{equation}

\begin{equation}
\begin{tabular}{|rrrr|rrrr|}
-1  &-1  &-1  &-1  &\ 0 &\ 0 &\ 0 &\ 0 \\
\ 0  &\ 0  & \ 0 &\ 0  &-1 &-1 &-1 &-1 \\
\hline
-1  &-1  &-1  &-1  &-1 &-1 &-1 &-1 \\
\end{tabular}
\end{equation}

\begin{equation}
\begin{tabular}{|rrrr|rrrr|}
\ 0  &\ 0  &\ 0  &\ 0  &\ 0 &\ 0 &\ 0 &\ 0 \\
\ 0  &\ 0  &\ 0  &\ 0  &\ 0 &\ 0 &\ 0 &\ 0 \\
\hline
\ 0  &\ 0  &\ 0  &\ 0  &\ 0 &\ 0 &\ 0 &\ 0 \\
\end{tabular}
\end{equation}

Observe that integration never stopped going through the full
period. However if integration is stopped at any point the pattern
will be inconsistent with the continuous case. Also we see that after a
full period the integration still needs to continue, as we have
returned to the original state. Hence there is no finite-length
loading using the input-side excitation method\footnote{This goes
counter to the intuition that one might be able to load the length of
the string once only using input-side integration, which is
incorrect.}. Again, off-center excitation follow the same pattern
without major differences.

\subsection{Output-side excitation}

Same excitation as before.

\begin{equation}
\begin{tabular}{|rrrr|rrrr|}
\ 0  &\ 0 &\ 0  & \ 0  &\ 0 &\ 0 &\ 0 &\ 0\\
\ 0  &\ 0 & \ 0 & \ 0  &\ 0 &\ 0 &\ 0 &\ 0\\
\hline
\ 0  &\ 0 & \ 0  & \ 0 &\ 0 &\ 0 &\ 0 &\ 0\\
\hline
\ 0  &\ 0 & \ 0  & \ 0 &\ 0 &\ 0 &\ 0 &\ 0\\
\end{tabular}
\end{equation}

\begin{equation}
\begin{tabular}{|rrrr|rrrr|}
\ 0  &\ 0  &\ 0  &\ 0  &\ 0 &\ 0 &\ 0 &\ 1 \\
\ 1  &\ 0  &\ 0  &\ 0  &\ 0 &\ 0 &\ 0 &\ 0 \\
\hline
\ 1  &\ 0  &\ 0  &\ 0  &\ 0 &\ 0 &\ 0 &\ 1 \\
\hline
\ 1  &\ 1  &\ 1  &\ 1  &\ 1 &\ 1 &\ 1 &\ 1 \\
\end{tabular}
\end{equation}

\begin{equation}
\begin{tabular}{|rrrr|rrrr|}
\ 0  &\ 0  &\ 0  &-1  &\ 0 &\ 0 &\ 0 &\ 0 \\
\ 0  &\ 0  & \ 0 &\ 0  &-1 &\ 0 &\ 0 &\ 0 \\
\hline
\ 0  &\ 0  &\ 0  &-1  &-1 &\ 0 &\ 0 &\ 0 \\
\hline
\ 0  &\ 0  &\ 0  &\ 0  &\ 0 &\ 0 &\ 0 &\ 0 \\
\end{tabular}
\end{equation}

\begin{equation}
\begin{tabular}{|rrrr|rrrr|}
\ 0  &\ 0  &\ 0  &\ 0  &\ 0 &\ 0 &\ 0 &-1 \\
-1  &\ 0  & \ 0 &\ 0  &\ 0 &\ 0 &\ 0 &\ 0 \\
\hline
-1  &\ 0  &\ 0  &\ 0  &\ 0 &\ 0 &\ 0 &-1 \\
\hline
-1  &-1  &-1  &-1  &-1 &-1 &-1 &-1 \\
\end{tabular}
\end{equation}

\begin{equation}
\begin{tabular}{|rrrr|rrrr|}
\ 0  &\ 0  &\ 0  &\ 1  &\ 0 &\ 0 &\ 0 &\ 0 \\
\ 0  &\ 0  &\ 0  &\ 0  &\ 1 &\ 0 &\ 0 &\ 0 \\
\hline
\ 0  &\ 0  &\ 0  &\ 1  &\ 1 &\ 0 &\ 0 &\ 0 \\
\hline
\ 0  &\ 0  &\ 0  &\ 0  &\ 0 &\ 0 &\ 0 &\ 0 \\
\end{tabular}
\end{equation}

Hence we see that the output-side integration yields the correct
pattern. The integration continues indefinitely by definition. However
compared to the input-side integrator we observe, that non-trivial
addition (adding zero) in the integrator happen twice per period for
one impulsive excitation, whereas the integrator on the input side
only needs to store the impulse and hence has no further non-trivial
additions after its initial loading. This means that under sparse
input conditions, {\em input-side integration is numerically
favorable}. More precisely, in the worst case an output-side
integrator will see indefinite non-trivial additions at every
time-step even for excitations of a maximum length of twice the
string. Input-side integrators will see non-trivial additions only at
every time-step when the excitation changes. If the excitation is
indefinitely non-trivial, the two methods are comparable with respect
to addition inaccuracies.

Output-side integration for lossless strings is impractical because
any numerical error in the addition will accumulate, though only
linearly, as the error is not fed back into the waveguide iteration.

Input-side integrators will feed numerical errors into the string,
yet again only linearly, as the content of the waveguide is not feed
back into the integrator.

This changes in case of non-linear coupling mechanisms and additional
care must be taken.

\section{Integration Methods and Neumann Boundary Conditions}

Next we discuss integration methods for strings with loose ends.

It is well known that this corresponds to reflections without sign
inversion at the boundary for displacement waves\cite{Graff91,Taylor96}.

\subsection{Heaviside Integration}

These are the respective results of the Heaviside Integrator for time
steps equivalent to half a string length. The excitation is at the midpoint:

\begin{equation}
\begin{tabular}{|rrrr|rrrr|}
1  & 1  & 1  & 1  &\ 0 &\ 0 &\ 0 &\ 0 \\
-1 & -1 & -1 & -1 &\ 0 &\ 0 &\ 0 &\ 0 \\
\hline
\ 0  &\ 0  &\ 0  &\ 0  &\ 0 &\ 0 &\ 0 &\ 0 \\
\end{tabular}
\end{equation}

\begin{equation}
\begin{tabular}{|rrrr|rrrr|}
 -1  & -1  & -1  & -1  &\ 1 &\ 1 &\ 1 &\ 1 \\
0  &\ 0  & \ 0 & \ 0 &\ 0 &\ 0 &\ 0 &\ 0 \\
\hline
-1  & -1  & -1  & -1  &\ 1 &\ 1 &\ 1 &\ 1 \\
\end{tabular}
\end{equation}

\begin{equation}
\begin{tabular}{|rrrr|rrrr|}
0  &\ 0  &\ 0  &\ 0  &\-1 &\-1 &\-1 &\-1 \\
0  &\ 0  & \ 0 & \ 0 & 1 & 1 & 1 & 1 \\
\hline
0  &\ 0  &\ 0  &\ 0  &\ 0 &\ 0 &\ 0 &\ 0 \\
\end{tabular}
\end{equation}

\begin{equation}
\begin{tabular}{|rrrr|rrrr|}
0  &\ 0  &\ 0  &\ 0  &\ 0 &\ 0 &\ 0 &\ 0 \\
 1  & 1  &  1 &  1 &-1 &-1 &-1 &-1 \\
\hline
 1  & 1  & 1  & 1  &-1 &-1 &-1 &-1 \\
\end{tabular}
\end{equation}

Hence we see that the Heaviside Integration does not yield the correct
accumulation of displacement as is seen in the continuous case
\cite{Essl04}. Why this simulation breaks down, remains to be
explored.

\subsection{Input-side Integration}

The excitation is placed in the middle of the string and constitutes
loading the result of a non-leaky integrator fed by an impulse to
equal parts left and right of the excitation-point into the respective
traveling waves.

\begin{equation}
\begin{tabular}{|rrrr|rrrr|}
0  &\ 0  &\ 0  &\ 0  &\ 0 &\ 0 &\ 0 &\ 0\\
\ 0  &\ 0 & \ 0 & \ 0  &\ 0 &\ 0 &\ 0 &\ 0 \\
\hline
\ 0  &\ 0  &\ 0  &\ 0  &\ 0 &\ 0 &\ 0 &\ 0 \\
\end{tabular}
\end{equation}

\begin{equation}
\begin{tabular}{|rrrr|rrrr|}
\ 0  &\ 0  &\ 0  &\ 0  &\ 1 &\ 1 &\ 1 &\ 1 \\
\ 1  &\ 1  &\ 1  &\ 1  &\ 0 &\ 0 &\ 0 &\ 0 \\
\hline
\ 1  &\ 1  &\ 1  &\ 1  &\ 1 &\ 1 &\ 1 &\ 1 \\
\end{tabular}
\end{equation}

\begin{equation}
\begin{tabular}{|rrrr|rrrr|}
 1  & 1  & 1  & 1  &\ 1 &\ 1 &\ 1 &\ 1 \\
\ 1  &\ 1  & \ 1 &\ 1  & 1 & 1 & 1 & 1 \\
\hline
\ 2  &\ 2  &\ 2  &\ 2  &\ 2 &\ 2 &\ 2 &\ 2 \\
\end{tabular}
\end{equation}

\begin{equation}
\begin{tabular}{|rrrr|rrrr|}
 1  & 1  & 1  & 1  &\ 2 &\ 2 &\ 2 &\ 2 \\
\ 2  &\ 2  & \ 2 &\ 2  & 1 & 1 & 1 & 1 \\
\hline
 3  & 3  & 3  & 3  & 3 & 3 & 3 & 3 \\
\end{tabular}
\end{equation}

\begin{equation}
\begin{tabular}{|rrrr|rrrr|}
\ 2  &\ 2  &\ 2  &\ 2  &\ 2 &\ 2 &\ 2 &\ 2 \\
\ 2  &\ 2  &\ 2  &\ 2  &\ 2 &\ 2 &\ 2 &\ 2 \\
\hline
\ 4  &\ 4  &\ 4  &\ 4  &\ 4 &\ 4 &\ 4 &\ 4 \\
\end{tabular}
\end{equation}

Observe that integration never stopped going through the full period
and again needs to continue. We observe the correct accumulation of
linear displacement \cite{Essl04}.

\subsection{Output-side excitation}

Same excitation as before.

\begin{equation}
\begin{tabular}{|rrrr|rrrr|}
\ 0  &\ 0 &\ 0  & \ 0  &\ 0 &\ 0 &\ 0 &\ 0\\
\ 0  &\ 0 & \ 0 & \ 0  &\ 0 &\ 0 &\ 0 &\ 0\\
\hline
\ 0  &\ 0 & \ 0  & \ 0 &\ 0 &\ 0 &\ 0 &\ 0\\
\hline
\ 0  &\ 0 & \ 0  & \ 0 &\ 0 &\ 0 &\ 0 &\ 0\\
\end{tabular}
\end{equation}

\begin{equation}
\begin{tabular}{|rrrr|rrrr|}
\ 0  &\ 0  &\ 0  &\ 0  &\ 0 &\ 0 &\ 0 &\ 1 \\
\ 1  &\ 0  &\ 0  &\ 0  &\ 0 &\ 0 &\ 0 &\ 0 \\
\hline
\ 1  &\ 0  &\ 0  &\ 0  &\ 0 &\ 0 &\ 0 &\ 1 \\
\hline
\ 1  &\ 1  &\ 1  &\ 1  &\ 1 &\ 1 &\ 1 &\ 1 \\
\end{tabular}
\end{equation}

\begin{equation}
\begin{tabular}{|rrrr|rrrr|}
\ 0  &\ 0  &\ 0  & 1  &\ 0 &\ 0 &\ 0 &\ 0 \\
\ 0  &\ 0  & \ 0 &\ 0  & 1 &\ 0 &\ 0 &\ 0 \\
\hline
\ 0  &\ 0  &\ 0  & 1  & 1 &\ 0 &\ 0 &\ 0 \\
\hline
\ 2  &\ 2  &\ 2  &\ 2  &\ 2 &\ 2 &\ 2 &\ 2 \\
\end{tabular}
\end{equation}

\begin{equation}
\begin{tabular}{|rrrr|rrrr|}
\ 0  &\ 0  &\ 0  &\ 0  &\ 0 &\ 0 &\ 0 & 1 \\
 1  &\ 0  & \ 0 &\ 0  &\ 0 &\ 0 &\ 0 &\ 0 \\
\hline
 3  &\ 3  &\ 3  &\ 3  &\ 3 &\ 3 &\ 3 & 3 \\
\hline
 3  & 3  & 3  & 3  & 3 & 3 & 3 & 3 \\
\end{tabular}
\end{equation}

\begin{equation}
\begin{tabular}{|rrrr|rrrr|}
\ 0  &\ 0  &\ 0  &\ 1  &\ 0 &\ 0 &\ 0 &\ 0 \\
\ 0  &\ 0  &\ 0  &\ 0  &\ 1 &\ 0 &\ 0 &\ 0 \\
\hline
\ 0  &\ 0  &\ 0  &\ 1  &\ 1 &\ 0 &\ 0 &\ 0 \\
\hline
\ 4  &\ 4  &\ 4  &\ 4  &\ 4 &\ 4 &\ 4 &\ 4 \\
\end{tabular}
\end{equation}

Hence we see that the output-side integration yields the correct
pattern\cite{Essl04}. The integration continues indefinitely by
definition.

\section{Damping and Loop Filters}

Finally I want to discuss the impact of filters, in particular the
most basic case of linear damping filters, corresponding to a gain
multiplication in the loop, on the schemes discussed. It is easy to see
that the Heaviside integration method is well-behaved with respect
to such loop filter. All the data is present and at least linear
damping will dissipate all information without trouble. The impact of
phase delay and its relationship to physically observable effects on
the waveform on the string is more complicated and should be
investigated separately. The same in fact holds for non-integrating
simulation, where the exact relationship between phase delay and
physically correct wave form is usually phenomenologically treated.

The output-side integration case is troubled by a linear gain loop
filter. Amplitude in the waveguide is dissipated, but prior states in
the output integrator may still have older higher amplitudes. Hence
subtracting them will leave an incorrect remainder in the integrator
proportional to $1-g$ where $g$ is a positive gain factor less than 1.
A potential solution is to introduce a matched leak to the
integrator. However, a precise match is critical to avoid numerical
inaccuracies.

The input-side integration case is difficult to damp with loop
filters, as the integrator will indefinitely feed input into the
loop. Hence the filters needs to be matched at the input to avoid this
problem.

Overall, of all the integration methods studied, only the {Heaviside
integration} method is numerically well-behaved, and easily adaptable
to loop filters as customary in waveguide synthesis. However it cannot
be used for a Neumann simulation in its current form. The other
methods have to be handled with care for they use non-leaky
integrators, which are numerically unstable.

\section{Computational cost}

The computational cost of the Heaviside integration method is
dependent on the excitation position. If the excitation position is at
the far end of the string the maximum integration length of twice the
string length $L$ (once for the each rail) is reached. If a choice of
rail direction is permissible in the particular implementation, this
can be reduced to one string length as the positive rail is chosen to
correspond to the shorter side of the string. This integration has to
be performed per impulsive velocity excitation $V$. Hence we get an
overall computational complexity $O(V\cdot L)$ and if $L$ is treated
as a constant $O(V)$. This is independent and in addition to the
complexity of the waveguide iteration. We denote by $O(WG)$ the
complexity of the waveguide computation accumulated over time
steps. Hence we get the total complexity of $max(O(V\cdot L),O(WG)$.

The computational cost of input-side integration is one non-trivial
loading per time step of the waveguide iteration. Additionally a
constant amount of operations are necessary for changes in the
integrator on non-trivial input. Hence the
complexity is $max(O(V),O(WG))$.

The computational cost of output-side integration depends on the
spatial distribution of the integration. Typically only one
observation point is of interest. Then one integration per time step
is necessary, independent of the output. The complexity is thus
$O(WG)$. If the full length of the string is required this becomes
$O(WG\cdot L)$.

We observe that the local, output-side integration is computationally
most efficient, while numerically least desirable. The Heaviside
integrator is never cheaper than the input-side integrator, the extra
cost depending on the length of the string and the position of the
excitation. Yet this is bought at desirable numerical properties and
easy of use.

\section{Difficulties with ``Localized Velocity Input''}

It is a repeated belief that traveling velocity waves can be
calculated from displacement waves by calculating their difference
instead of their sum \cite{Smith97,Smith04}. It is considered a form
of localized velocity excitation. Difficulties with this belief has in
essence already been addressed by the author
elsewhere\cite{Essl04}. Here I would like to discuss this difficulty
based on the given examples.

Assume that one wants to simulate a local velocity impulse and it's
effect on displacement. Using the above prescription, one might
implement the following algorithm: [Draft Note: Needs introduction to
notation. This relates to \cite{Essl04,Smith04}] $y^+_{n,m}=1$ and
$y^-_{n,m}=-1$, which is the difference of the standard displacement
impulse $y^+_{n,m}=1$ and $y^-_{n,m}=1$, ignoring potential needs for
rescaling. Initially there is no displacement in accordance with
expectations for a velocity excitation. However, as time evolves on
sees an isolated negative impulse traveling left and an isolated
positive impulse traveling right (see also Eq. (45)
\cite{Essl04}). However if we excite the lossless wave equation with a
velocity impulse, we get a different picture. We get a spreading
square pulse \cite{Graff91,Essl04,Smith04}.

Clearly the simple use of the relationship between displacement and
velocity waves in waveguides gives an incorrect result.

The interpretation of difference of displacement waves can also be
seen in the simulations provided here. Observe equations
(\ref{eq:h1}-\ref{eq:h3}).

We see that these pictures do not violate the naive interpretation of
the relation of velocity to displacement waves. Indeed we have zero
displacement everywhere and the data present make the displacement
zero by using the difference.

However, it does give us a clue as to the difficulty in using the
interpretation that lead us to the naive approach in the first
place. If we accept that the difference between rails gives velocity,
and we observe that a step function has been loaded into the
waveguide, we have to conclude that the waveguide contains an
semi-infinitely extended velocity. Rather than the string moving upward on the whole of the semi-infinite half-line, it starts to spread only from the initial excitation point.

Clearly there is a difficulty in using the naive interpretation.

Clearly the difference between the two traveling wave rails cannot be
velocity. There is another reason for this, which is
dimensionality. The sum and the difference of two variable of the same
dimensionality will stay of that dimensionality. The sum and
difference of displacement stays a displacement. So one has to not
only take the difference to get displacement waves, one has to also
integrate inertial velocities to make sure they have the correct
dimensionality. This integration creates the step functions that we
observe above. This is indeed peculiar, as step functions are non-zero
out into infinity, hence are non-local. It has in fact been pointed
out that variables in waveguides in some constellations have this
non-local property\cite{Karjalainen03,KE03}. The string, like any
mechanical system, requires displacement and velocity for full
specification \cite{Smith97} so we observe that one of them has
a non-local property when compared to the other.

Hence ``localized velocity excitations'' should be considered a
displacement excitation in terms of dimensionality of the quantities
involved. It also shares the time-symmetric properties of displacement
excitations (see eq. (49) in \cite{Essl04}).

This indicates that one can in fact not so readily go from
displacement to velocity. One is non-local to the other and the
conversion is not only difference but also integration or
differentiation, see \cite{Smith97}.

\section{Conclusions}

We discussed three integration method for velocity excitations in
displacement waveguide simulations. They differ in terms of numerical
properties, relation to loop filters, computational cost and
generality. For most situations the Heaviside integration algorithm is
most desirable, except for strings with loose ends, when this methods
is incorrect. Of the remaining methods, input-side integration is
generally more desirable, than output side integration, for numerical
reasons and for the impact on loop filters.

We also discussed the difficulties with localized velocity excitations
as difference of displacement waves and the impact of the change of
wave variables on the loop filter in use.

Localized velocity excitations will generally yields results different
from the wave equation. Loop filters not designed with the integrating
behavior of displacement waves in mind, may inaccurately model the
desired behavior. Hence sufficient care must be taken.

It is worthwhile to note, that the excitation algorithms presented
here don't constitute a complete excitation description with respect
to the wave equation. In general both displacement and velocity waves
are present at the same time and hence excitations of the both types
can occur in any linear mixture\cite{Essl04}.

\begin{acknowledgments}
I am grateful for discussions with Matti Karjalainen, Cumhur Erkut and
for comments of two anonymous referees of an earlier manuscript in
review\cite{Essl04} which all contributed to my thinking on the
subject of this manuscript.
\end{acknowledgments}


\end{document}